\documentclass[aps,showpacs,twocolumn,preprintnumbers,amsmath,amssymb,pra]{revtex4-1}

\usepackage{graphicx}
\usepackage{dcolumn}
\usepackage{bm}
\usepackage{color}

\begin{document}

\title{Emission of dispersive waves from a train of dark solitons in optical fibers}

\author{T. Marest,$^{1}$ C. Mas Arab\'{i},$^1$ M. Conforti,$^1$ A. Mussot,$^1$ C. Mili\'{a}n,$^2$ D.V. Skryabin,$^{3,4}$ and A. Kudlinski$^{1,}$}
\email{Corresponding author: alexandre.kudlinski@univ-lille1.fr}

\affiliation{
$^1$Univ. Lille, CNRS, UMR 8523 - PhLAM - Physique des Lasers Atomes et Mol\'{e}cules, F-59000 Lille, France\\
$^2$Centre de Physique Th\'{e}orique, \'{E}cole Polytechnique, CNRS, Universit\'{e} Paris-Saclay, F-91128 Palaiseau, France\\
$^3$Department of Physics, University of Bath, Bath, United Kingdom\\
$^4$ITMO University, Kronverksky Avenue 49, St. Petersburg 197101, Russian Federation}

\begin{abstract}
We report the experimental observation of multiple dispersive waves emitted in the anomalous dispersion region of an optical fiber from a train of dark solitons. Each individual dispersive wave can be associated to one particular dark soliton of the train, using phase-matching arguments involving higher-order dispersion and soliton velocity. For a large number of dark solitons ($>10$), we observe the formation of a continuum associated with the efficient emission of dispersive waves.
\end{abstract}
													

\maketitle

A bright temporal soliton propagating in the anomalous dispersion region of an optical fiber and in the neighbourhood of the zero-dispersion wavelength (ZDW) is known to emit a resonant radiation across the ZDW (in normal dispersion region) \cite{Wai1986,Wai1987}. The physical picture associated to this process is as follows: a resonant frequency belonging to a part of the soliton spectral tail overlapping with the normal dispersion region is radiated in the form of a dispersive wave (DW), satisfying a phase-matching condition involving higher-order dispersion \cite{Akhmediev1995}. This process has been studied extensively from theoretical and experimental points of views, in particular in the context of supercontinuum generation in photonic crystal fibers (PCFs) in which it plays a crucial role \cite{Cristiani2004,Dudley2006,Skryabin2010}. Recently, this process has been shown to occur with non-solitonic pulses propagating in normal dispersion region, thus exciting resonant DWs in the anomalous dispersion region \cite{Webb2013}. The physical explanation for this observation has been provided soon after in terms of phase-matching between the shock-wave front produced from wave breaking in normally dispersive fibers and the resonant DW \cite{Conforti2013}. In this case, phase-matching arguments have to account for the group velocity of the shock front \cite{Conforti2013,Conforti2014pra}, whose effect, in the case of anomalous dispersion and bright solitons, is usually negligible with respect to the one of higher-order dispersion.

In fact, a few early theoretical studies revealed that dark solitons propagating in normal dispersion region may also radiate resonant DWs under the action of third-order dispersion \cite{Karpman1993,Afanasjev1996}, which can even initiate the formation of a continuum \cite{Milian2009}. Very recently, the interaction between a dark soliton and an external dispersive wave has been numerically studied in \cite{Oreshnikov2015}. Besides these few theoretical studies, the generation of DWs from dark solitons has not been much investigated and has never been observed experimentally.

In this work, we report the first experimental observation of the emission of DWs from a train of dark solitons. Our numerical analysis reveals that each emitted DW can be identified to one particular dark soliton of the train using a phase-matching relation involving higher-order dispersion and dark soliton velocity. Finally,  following numerical predictions from \cite{Milian2009}, we provide an experimental demonstration of the formation of a spectral continuum formed by DWs emitted by multiple dark solitons.


A simple method to generate a train of dark soliton consists in launching two delayed co-propagating pulses in a normally dispersive optical fiber \cite{Rothenberg1991,Rothenberg1992}. At the initial stage of propagation, the two pulses temporally broaden and acquire a linear chirp due to dispersion and self-phase modulation. Once they overlap in time, they interfere forming a pulse with sinusoidally modulated intensity. With further propagation, this temporal modulation nonlinearly reshapes producing a train of isolated dark pulses, asymptotically approaching fundamental dark solitons \cite{Rothenberg1991}. The dark pulses behave as quasi-dark solitons, but they are not fundamental dark solitons in the strict sense. In fact, they either accelerate or decelerate on a variable background and their darkness (or rather greyness) changes with propagation. In the following, we chose to term them dark solitons anyway, for the sake of simplicity and clarity.

\begin{figure}[t!]
\centering
\includegraphics[width=\linewidth]{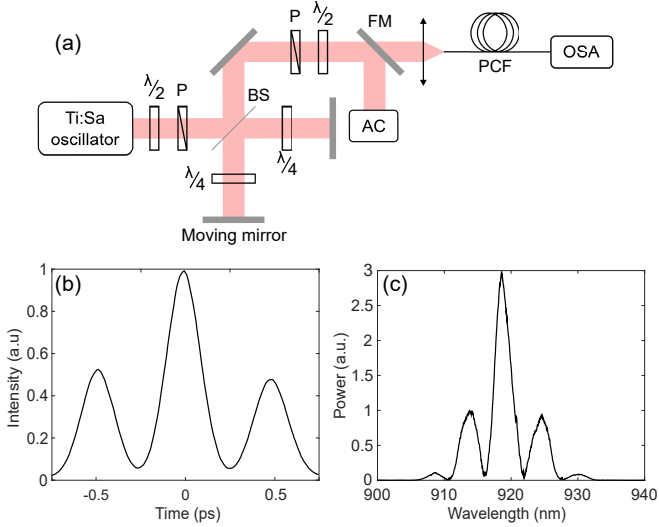}
\caption{(a) Experimental setup. P : polarizer; BS : beam splitter; FM : flipping mirror; OSA : optical spectrum analyser; AC : autocorrelator. (b) Typical autocorrelation trace of the two input delayed pulses used to form the dark soliton train. (c) Corresponding input spectrum (linear scale).}
\label{fig:setup}
\end{figure}

Figure \ref{fig:setup} shows a scheme of the experimental setup we have implemented to use this technique. A Ti:Sa oscillator delivers Gaussian pulses with a full width at half-maximum (FWHM) duration of 140 fs tunable around 800 nm at 80 MHz. The pulses pass through a variable attenuator made of a half-wave plate and a polarizer and they are directed into an unbalanced Michelson interferometer producing a pair of delayed pulses, with a delay adjustable by moving one of the mirrors with a micrometer screw. In each arm, a quarter-wave plate is used to adjust the polarization state resulting in the power control of each pulse after a polarizer placed at the interferometer output. The final half-wave plate fixes the polarization state of the two pulses to a neutral axis of the fiber. The pulse duration and delay at the fiber input are measured with a noncollinear autocorrelator and the corresponding spectrum is measured with an optical spectrum analyzer. Figures \ref{fig:setup}(b) and (c) show respectively the typical autocorrelation trace and spectrum of the input pulse pair used for the generation of the dark soliton train in the fiber. In this example, the delay was about 500 fs.

The next step of our study is to design a fiber suitable for (i) the generation of a train of dark solitons from two delayed 140 fs pulses and (ii) the subsequent emission of DWs from this dark soliton train. The fiber must be normally dispersive to allow the formation of dark solitons and its ZDW must be close to the emission wavelength of our Ti:Sa oscillator to allow DW emission. We therefore used a PCF whose guiding properties were calculated using the empirical model from Ref.~\cite{Saitoh2005}. The optimal pump conditions (wavelength, power, delay) and fiber properties (length, dispersion, nonlinearity) were determined from numerical simulations using the following generalized nonlinear Schr\"{o}dinger equation (GNSLE) :
\begin{equation}
\small{i\partial_zA+D(i\partial_t)A+\gamma \bigg(1+i\tau_{s}\partial_t\bigg)\bigg(A\int R(t')|A(t-t')|^2 dt' \bigg) = 0}
\label{GNLSE}
\end{equation}
where the dispersion operator $D(i\partial_t)=\sum_{{n \ge 2}} \frac{\beta_n}{n!} \,( i\partial_t )^n$ takes into account the full dispersion profile of the fiber and  $\gamma$ is the nonlinear parameter. $R(t)=(1-f_R)\delta(t)+f_R h_R(t)$ includes both Kerr and Raman responses ($f_R=0.18$) \cite{Hollenbeck2002} and $\tau_{s} \approx 1/\omega_{p}$ is the self-steepening term. $D(i\partial_t)$ is expanded around $\omega_p$, the pump carrier frequency, and $t$ is the retarded time in the frame travelling at the group velocity $V_g=V_g(\omega_p)=\beta_1^{-1}$ of the pump pulses. From these preliminary numerical simulations, we chose to fix the pump wavelength to 920 nm and we used a PCF with a hole-to-hole spacing $\Lambda$ of 2.88 µm and a relative hole diameter $d/\Lambda$ of 0.62. The corresponding ZDW is located near 970 nm, and the nonlinear parameter equals 22 W$^{-1}$.km$^{-1}$ at 920 nm.

\begin{figure}[t!]
\centering
\includegraphics[width=\linewidth]{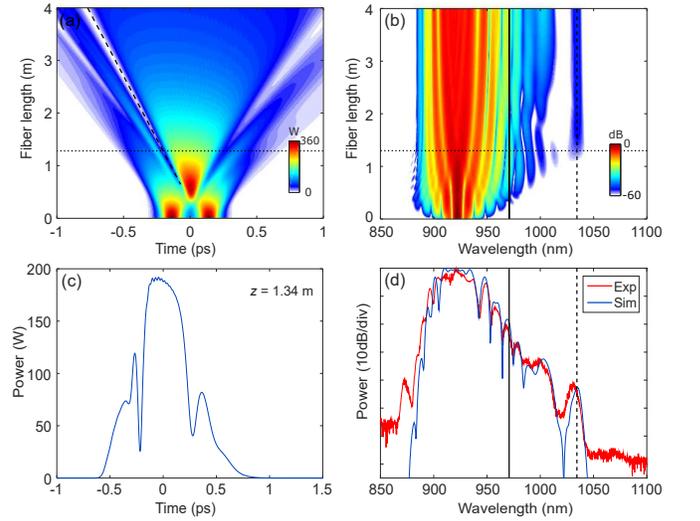}
\caption{DW generation by a single dark soliton.  (a) Simulated temporal evolution versus fiber length. The dashed line represent the calculated soliton velocity $V_s$. (b) Simulated spectral dynamics. The vertical solid and dashed lines correspond respectively to the ZDW and the DW wavelength calculated using Eq. (\ref{disp_rel}). In (a) and (b), the fiber length at which the DW is emitted ($z = 1.34$ m) is represented by dotted horizontal lines. (c) Simulated temporal evolution at $z = 1.34$ m. (d) Simulated (blue line) and measured (red line) spectra at the output of a 4 m-long PCF. Initial pulse delay: 280 fs.}
\label{fig:1wave}
\end{figure}
Figure \ref{fig:1wave} summarizes the results obtained in this configuration, for a pump peak power of 360 W for each pulse and a delay of 280 fs between them. The temporal evolution versus fiber length, plotted in Fig. \ref{fig:1wave}(a), shows the formation of two dark solitons after a typical propagation distance of 1 m. In the corresponding spectral evolution shown in Fig. \ref{fig:1wave}(b), the spectrum progressively broadens across the ZDW (represented by the vertical solid line) and a radiation, possibly a DW, is generated at 1034 nm in the anomalous dispersion region starting from a fiber length of 1.34 m. A closer look at the temporal profile at this particular propagation distance is provided in Fig. \ref{fig:1wave}(c), where two dark solitons appear on both edges of the pulse. It can be observed that the dark soliton located on the pulse leading edge is much sharper than the one on the trailing edge. Our aim now is thus to identify from which of these two solitons originates the 1034 nm radiation.

To perform this analysis, we use a perturbative approach. Let us write the field as $A(z,\tau)=(F(\tau)+g(\tau,z))\text{e}^{ik_{NL}z}$, where $F(\tau)$ is assumed to be a dark soliton, $g(\tau,z)$ represents a weak DW, $k_{NL}=\gamma|A_0|^2$ is a constant proportional to the dark soliton background $A_0$, and $\tau=t-z/V_{s}$ is time in a reference frame such as the soliton is at rest. By taking $f_R=0$ and $\tau_s=0$ in Eq. (\ref{GNLSE}) and substituting $A(z,\tau)$ we obtain :
\begin{align}
i \partial_z g+\tilde{D} (i\partial_\tau)g - k_{NL}g + 2|F|^2g+F^2g^*=  \nonumber \\  =-\left[D( \partial_\tau)-\frac{\beta_2}{2}(i\partial_\tau)^2\right]F
\label{radiation}
\end{align}
where we have defined $\tilde{D}(i\partial_\tau)=-iV_s^{-1}\partial_{\tau}+D(i\partial_\tau)$. If we write the field $g(z,\tau)=a(z)\text{e}^{i(\lambda z-\delta \tau)}+b^*(z)\text{e}^{-i(\lambda z-\delta \tau)}$ and follow the standard procedure described, for instance, in \cite{Karpman1993,Conforti2014}, we get :
\begin{equation}
\lambda_{\pm}=\frac{d_{odd}\pm\sqrt{d_{even}^2+4k_{NL}d_{even}}}{2}=0,
\end{equation}
where we have defined $d_{odd}=\tilde{D}(\delta)-\tilde{D}(-\delta)$ and $d_{even}=\tilde{D}(\delta)+\tilde{D}(-\delta)$, and where $\lambda = 0$ was imposed to set $g(z,t)$ in phase-matching with the soliton. In order to simplify this equation, we assume $\tilde{D}(\delta)+\tilde{D}(-\delta)\gg$ $k_{NL}$, allowing us to perform an expansion of the square root, which gives the following phase-matching relation :
\begin{equation}
D(\delta)-\delta /V_{s}+k_{NL}=0.
\label{disp_rel}
\end{equation}
where $V_s$ is the soliton velocity relatively to the pump, meaning that $V_s > 0$ (resp. $V_s < 0$) for solitons travelling slower (resp. faster) than the pump. Note that in the case of bright pulses on a vanishing background, the $k_{NL}$ term in Eq. \ref{disp_rel} would be preceded by a minus sign \cite{Conforti2014pra}.

Using the numerical simulation of the temporal evolution versus fiber length [Fig. \ref{fig:1wave}(a)], we can measure the velocity $V_{s}$ and the amplitude of the soliton at the propagation distance (represented with a horizontal dotted line) where the DW is emitted. The black dashed line in Fig. \ref{fig:1wave}(a) illustrates how the soliton velocity $V_{s}$ is determined in the $(t,z)$ plane at $z = 1.34$ m from this numerical temporal map. Then, reporting this value of $V_s$ in Eq. (\ref{disp_rel}), we can calculate the wavelength of the DW emitted by each soliton of the train. A first analysis points out that the sign of $V_s$ changes dramatically the phase-matched DW wavelength. For the slowest soliton ($V_{s}>0$), the phase-matching relation gives a DW wavelength of 1120 nm, which is much further from the ZDW than the one emitted from the fastest one ($V_{s}<0$) located at 1034 nm. Furthermore, because the third-order dispersion term $\beta_{3}$ is positive here, the fastest soliton has a Raman frequency shift closer to the ZDW \cite{Milian2009} and so it is more susceptible to emit radiation with a good efficiency. The DW wavelength of 1034 nm calculated from Eq. (\ref{disp_rel}) is represented by the vertical dashed line in Fig. \ref{fig:1wave}(b), which turns out to be in excellent agreement with the radiation observed in simulations. Using the experimental setup of Fig. \ref{fig:setup} and using a PCF fabricated with exactly the same parameters as described above, we compare in Fig. \ref{fig:1wave}(d) the experimental and numerical spectra at the output of a 4 m-long PCF, for an input pulse delay of 280 fs. Both spectra are in excellent agreement, and show the evidence of DW emission at 1034 nm, in excellent agreement with the solution of Eq. (\ref{disp_rel}) represented by the dashed vertical line. This experiment constitutes the first observation of DW emission from a dark soliton.

\begin{figure}[t!]
\centering
\includegraphics[width=\linewidth]{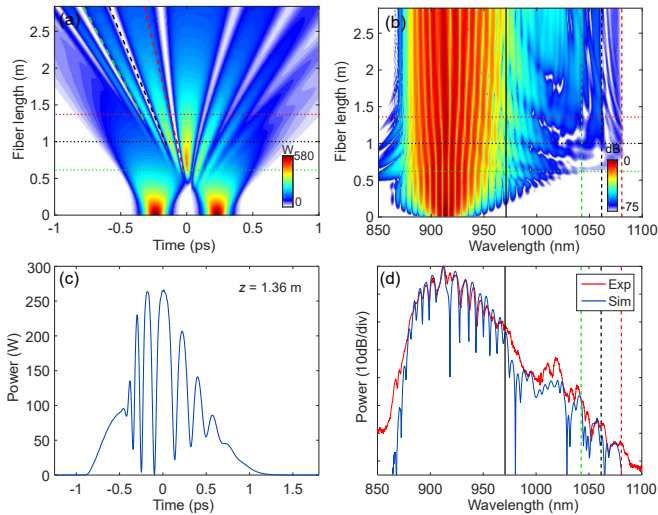}
\caption{Generation of DWs from multiple dark solitons. (a) Simulated temporal evolution versus fiber length. The dashed lines correspond to the velocity of each soliton emitting a DW. (b) Corresponding spectral evolution versus length. The vertical solid and dashed lines correspond respectively to the ZDW and the DW wavelengths calculated using Eq. (\ref{disp_rel}) with the corresponding soliton velocity [of same color as in (a)]. In (a) and (b), the fiber lengths at which DWs are emitted ($z =$ 1.36, 1 and 0.62 m) are represented by dotted horizontal lines. (c) Simulated temporal evolution at $z = 1.36$ m, i.e. where the first DW is emitted. (d) Simulated (blue line) and measured (red line) spectra at the output of a 2.8 m-long PCF. Initial pulse delay: 465 fs.}
\label{fig:3wave}
\end{figure}

In a second configuration, we aimed at studying the DW emission process from a pulse containing a larger number of dark solitons. This can be done by simply increasing the delay between the two input pulses \cite{Rothenberg1991}. Figure \ref{fig:3wave} summarizes the numerical and experimental results obtained for an input delay of 465 fs. All other parameters and conditions are similar to the configuration of Fig. \ref{fig:1wave}, except the fiber length which was reduced to 2.8 m and the peak power which was increased to 590 W. The temporal map plotted in Fig. \ref{fig:3wave}(a) shows that six dark solitons are formed within the first meter of propagation.  The corresponding spectral map shown in Fig. \ref{fig:3wave}(b) exhibits three main radiations in the anomalous dispersion region. The first one, generated after a propagation length of 1.36 m, is located at 1038 nm, the second one generated after 1 m is located at 1058 nm, and the third one generated after 0.62 m is located at 1076 nm. The temporal profile simulated at 1.36 m (i.e. where the first DW is emitted) is represented in Fig. \ref{fig:3wave}(c). It shows that the three dark solitons located on the leading edge (with negative velocities) are much sharper and shorter than the three ones of the trailing edge (having positive velocities). As we did in the previous configuration, we calculated the velocity $V_s$ of each soliton from the $(t,z)$ map (illustrated by dashed lines in Fig. \ref{fig:3wave}(a) for the three shortest dark solitons) and we calculated the corresponding phase-matched DW wavelength from Eq. (\ref{disp_rel}). They are represented in Fig. \ref{fig:3wave}(b) with the corresponding color, and show good qualitative agreement with the observed radiations in simulations. Finally, Fig. \ref{fig:3wave}(d) shows the experimental output spectrum in red solid line, which is in excellent agreement with the simulated one (in blue) and with solutions of the phase-matching relation (Eq. (\ref{disp_rel})) for the three fastest solitons (vertical dashed lines). The slight discrepancy between simulations/experiments and solutions of Eq. (\ref{disp_rel}) is due to the uncertainty in determining the amplitude and velocity of the dark solitons, as explained above. These results highlight the generation of three DWs originating from the three different dark solitons which can be identified through their group velocity.

\begin{figure}[t!]
\centering
\includegraphics[width=\linewidth]{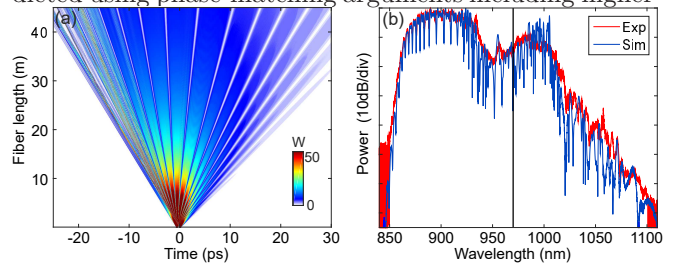}
\caption{Continuum generation from dark solitons. (a) Simulated temporal evolution at the output fiber. (b) Simulated (blue line) and measured (red line) output spectra for a fiber length of 45 m. Initial pulse delay: 624 fs.}
\label{fig:SC}
\end{figure}

The experimental results above and numerical studies from \cite{Milian2009} show that increasing the number of dark solitons in the train allows to generate multiple DWs, which enhances the overall spectral broadening into the anomalous dispersion region. Eventually, this may form a continuum, as shown numerically in \cite{Milian2009}. In order to observe this experimentally, we slightly increased the initial delay to 624 fs (in order to increase the number of dark solitons) and used a longer PCF of 45 m. In this case, the PCF design was slightly different from the previous one, although very close. The ZDW is 960 nm and the nonlinear parameter is 23 W$^{-1}$.km$^{-1}$ at 900 nm (the whole PCF properties can be easily calculated using \cite{Saitoh2005}, with $d/\Lambda = 0.57$ and $d$ = 2.64 µm). With this PCF, the optimal pump wavelength was 900 nm. Figure \ref{fig:SC}(a) shows the temporal dynamics simulated with a pump peak power of 1 kW for each pulse, highlighting the generation of multiple ($>10$) dark solitons. In addition, distortions appear at the leading edge of the pulse, possibly due to the formation of a shock wave front, which can also contribute to the emission of a DW \cite{Conforti2013,Conforti2014pra}. The experimental and numerical output spectra [red and blue lines in Fig. \ref{fig:SC}(b), respectively] are in excellent agreement, and show the formation of a continuum due to multiple DWs emitted by dark solitons, following the process analyzed in details in \cite{Milian2009}. In particular the experimental result of Fig. \ref{fig:SC}(b) reproduce qualitatively the results of Fig. 5(a) from \cite{Milian2009}, in which Raman scattering is shown to enhance the efficiency of DW emission from dark solitons.

\section*{Conclusion}

In summary, we have experimentally observed for the first time the emission of dispersive waves from dark solitons. This was done by forming a train of dark solitons from the nonlinear collision of two identical and delayed pulses propagating in a normally dispersive fiber. The shortest and sharpest dark solitons of the train emit dispersive waves whose wavelength can be accurately predicted using phase-matching arguments including higher-order dispersion terms and soliton velocity. Increasing the number of dark solitons in the pulse results in the formation of a continuum, following previous numerical studies of \cite{Milian2009}.

These experimental results constitute a remarkable illustration of the generalization of the dispersive wave emission process to pulses propagating in the normal dispersion region \cite{Webb2013}.

\section*{Acknowledgement}

This work was partly supported by IRCICA, CNRS, USR 3380, by the ANR TOPWAVE (ANR-13-JS04-0004) and NoAWE (ANR-14-ACHN-0014) projects, by the "Fonds Europ\'{e}en de D\'{e}veloppement Economique R\'{e}gional", the Labex CEMPI (ANR-11-LABX-0007) and Equipex FLUX (ANR-11-EQPX-0017) through the "Programme Investissements d'Avenir". We also acknowledge funding from the CNRS and RFBR through the Russian-French PRC program. D.V.S. acknowledges support through the ITMO visiting professorship scheme.

\end{document}